\documentstyle[11pt,a4,amssymb,epsf]{article}

\newcommand{\be}{\begin{equation}}
 \newcommand{\ee}{\end{equation}} \newcommand{\bea}{\begin{eqnarray}}
 \newcommand{\eea}{\end{eqnarray}}

  \catcode`\@=11 \@addtoreset{equation}{section}
  \catcode`\@=11

\begin{document}

\begin{titlepage}

\begin{flushright} {\tt FTUV/98-38\\ IFIC/98-39 \\ SU-ITP/98-34 \\
gr-qc/9805082}
 \end{flushright}

\vfill

\begin{center}

{\bf{Critical energy flux and mass in solvable theories \\ of 2d dilaton gravity
}}

\bigskip \bigskip A. Fabbri \footnote{Supported by an INFN fellowship. 
e-mail: \sc{ afabbri1@leland.stanford.edu}}
\bigskip

\begin{center} Department of Physics,  
	Stanford University, \\ 
      Stanford, CA, 94305-4060, USA\\
 \bigskip \bigskip 
and J.
Navarro-Salas\footnote{Work partially supported by the 
Comisi\'on Interministerial de Ciencia y Tecnolog\'{\i}a and DGICYT.
e-mail: \sc{ jnavarro@lie.ific.uv.es}}

\bigskip

Departamento de F\'{\i}sica Te\'orica and\\ IFIC, Centro Mixto
	Universidad de Valencia-CSIC.\\ Facultad de F\'{\i}sica, Universidad de
	Valencia,\\ Burjassot-46100, Valencia, Spain.\\

\end{center} \bigskip \today \bigskip

\end{center}


\begin{center} {\bf Abstract} \end{center}
In this paper we address the issue of determining the semiclassical 
threshold for black hole formation in the context of a one-parameter 
family of theories which continuously interpolates between the RST
and BPP models. 
We find that the results depend significantly on the initial static 
configuration of the spacetime geometry before the influx of matter is
turned on. In some cases there is a critical energy density, given by 
the Hawking rate of evaporation, as well as a critical mass $m_{cr}$ 
(eventually vanishing). In others there is neither $m_{cr}$ nor a
critical flux.

\vfill \end{titlepage} \newpage

\section{Introduction} 
Black holes are among the most fascinating and interesting objects in modern
theoretical physics.\ Discovered in the context of general relativity,
their understanding from the point of view of the quantum theory is one of
the 
essential ingredients in the search of a unified theory of all fundamental interactions.
\\
Classically black holes are ``simple'' objects, i.e. as their name suggests
they absorb any kind of matter but since light itself gets trapped in their
gravitational field they are invisible to any external observer. 
This view has however been drastically modified by quantum considerations.
The basic process can be understood, heuristically, by considering loops
of virtual particles 
close to the
event horizon; the gravitational field of the hole is capable to capture one 
partner (provided its energy is negative) leaving the other free to reach 
infinity. Hawking \cite{Hawking} has shown, in fact,
that they rather behave
as hot bodies with temperature $T_H={k_+\over2\pi}$ \footnote{Here and 
throughout the paper 
we will consider units where $\hbar=G=c=1$}, where $k_+$ is the surface gravity
at the event horizon. \\
In this paper we will consider the aspect of the formation of black holes 
in a simplified context, namely two dimensional dilaton gravity.
In the classical theory in 1+1 dimensions collapse of matter (in the form of 
conformally coupled scalar fields) always forms a stable black hole, no 
matter the amount of total incoming energy $M$.\footnote{In four dimensions, however,
there is a classical threshold for black hole formation, see 
 \cite{Choptuik} .}
The discovery of exactly solvable models at the semiclassical level
\cite{Bilal.Callan}, where the backreaction of the Hawking radiation 
on the background geometry can be analytically evaluated, has been very useful
for understanding many features of quantum black hole physics. \\
In the present context we will consider a one-parameter ($a$) family
of models introduced in \cite{Cruz.Salas} given by the action $S=S_{cl} +S_q$, where 
 \be
  S_{cl} ={1\over2\pi}\int d^2
 x\sqrt{-g}\left[e^{-2\phi}\left(R+4\left(\nabla\phi\right
 )^2+4\lambda^2\right)-{1\over2}\sum_{i=1}^N\left(\nabla f_i\right)^2 
\right]
  \> \label{i}
  \ee
and
\be
  S_q ={1\over2\pi}\int d^2
 x\sqrt{-g}\left[
-{N\over 48} R\square^{-1}R + {N\over 12}\left(1-2a\right)
 \left(\nabla\phi\right)^2+\left(a-1\right)\phi R
\right] 
  \>.\label{ii}
  \ee
For $a=1/2$ we recover the RST model \cite{RST} and when $a=0$ the one given by BPP
\cite{BPP}.
The classical limit of these theories, i.e. $S_{cl}$, is the CGHS model \cite{CGHS},
which describes low-energy excitations along the infinite throat of extremal 
(magnetic) stringy black holes in four dimensions.  
Its general static solution is simply expressed 
in terms of a mass parameter $M$.  When $M>0$ it is a black hole and has the same causal
structure of the Schwarzschild solution; the case $M=0$ is the well known linear 
dilaton vacuum and, finally, for $M<0$ the spacetime geometry exhibits a 
naked timelike singularity. \\
In the semiclassical regime (which, we remind, makes sense as an approximation to the
full quantum theory only for $N\to \infty$ , $Ne^{2\phi}$ fixed) it turns out 
that by requiring the absence of radiation at infinity Minkowski spacetime is no more 
solution to the equations of motion unless $a=1/2$ (i.e. RST).
For different values of $a$ the ``ground state'' of the theory is a nonflat geometry
asymptotically minkowskian (as $e^{2\phi}\to 0$)  and, in the strong coupling region,
with generically a regular timelike boundary at a finite proper distance from any 
other point (for $a=0$ it becomes, instead, an infinite throat
and the spacetime is geodesically complete). These solutions also
represent the end-point of the Hawking evaporation process. \\
There are however other solutions, obtained by imposing reflecting boundary
conditions along some timelike surface in the strong coupling region, which 
can also be considered regular from the point of view of the semiclassical theory.
We will use all such configurations as possible initial states for the gravitational
collapse process that we will investigate. Starting with the simple case of an incoming
shock-wave (section 4), we will then consider a constant energy density flux (section 5)
and show finally, in section 6, that the results obtained have a rather general
validity and apply for all types of collapsing null matter.  
            
 
\section {The CGHS model: classical solutions}
 In this section we will recall briefly the form of the classical solutions.
The CGHS theory is given by the action $S_{cl}$ of eq. (\ref{i}), where
$R$ is the 2d 
Ricci scalar, $\phi$ the dilaton field, $\lambda^2$ the cosmological constant and $f_i$
represent $N$ massless conformally coupled scalar fields. \\
Choosing conformal frame $ds^2=-e^{2\rho}dx^+dx^-$ the equations of motion of this theory
obtained by variation with respect to the metric are
\be
e^{-2\phi}(4\partial_{\pm}\rho\partial_{\pm}\phi-2\partial_{\pm}^2\phi)
+\sum_{i=1}^N {1\over2}\partial_{\pm}f_i\partial_{\pm}f_i=0
\>,
 \label{ec}
 \ee
\be
e^{-2\phi}(2\partial_+\partial_-\phi -4\partial_+\phi\partial_-\phi-\lambda^2e^{2\rho})=0
\>.
 \label{ec}
 \ee
Variation of the dilaton and the matter fields gives
\be 
-4\partial_+\partial_-\phi +4\partial_+\phi\partial_-\phi +2\partial_+\partial_-\rho
+\lambda^2 e^{2\rho}=0
\>,
 \label{ed}
 \ee
\be
\partial_+\partial_- f_i =0
\>.
 \label{em}
 \ee
It is possible to fix the residual diffeomorphism invariance (i.e. the
transformations $x^{\pm}\to x^{' \pm}(x^{\pm})$ that preserve the
conformal frame) 
and impose the Kruskal gauge choice
  \be
\rho=\phi\>\label{iii}
\ee 
for which the static solutions to the equations of motion take the simple form
 \be
e^{-2\phi}=e^{-2\rho}={M\over\lambda} - \lambda^2 x^+x^-\>.\label{iv}
\ee  
The parameter $M$ is identified with the ADM mass. \\
The solutions with $M>0$ represent black holes: they are characterized by a spacelike
curvature singularity located at $x^+x^-={M\over\lambda^3}$, event horizons at
$x^{\pm}=0$ and are asymptotically minkowskian as $x^+x^-\to\infty$. \\
The case $M=0$ is the linear dilaton vacuum. This is easily seen by transforming 
to coordinates $\sigma^{\pm}$ such that 
$\pm\lambda x^{\pm}=e^{\pm\lambda\sigma^{\pm}}$,
where 
 \be
ds^2=-d\sigma^+d\sigma^- \ ,\  \phi=-\lambda \sigma
\>\label{v}
\ee  
and 
$\sigma \equiv {({\sigma^+ -\sigma^-})\over 2}$.\\
Finally, when $M<0$ there is a timelike singularity at $x^+x^-=-{|M|\over\lambda^3}$.
By cosmic censorship arguments this solution should be excluded from the physical
spectrum. However, we will see in the next section that
we can nonetheless introduce semiclassical configurations which reduce, in the classical
limit, to these solutions. This simple fact
will be important for the discussion 
of our results.


 \section{Semiclassical static solutions of the RST-BPP models and spacetime structure}
The solvability of the semiclassical theory $S_{cl} + S_q$, given in  (\ref{i}) and 
 (\ref{ii}), is essentially due to the fact that provided we perform the field
redefinitions   
 \be
 \Omega={N\over 12}a\phi\ + e^{-2\phi}\>,\label{vi}
 \ee
 \be
 \chi={N\over 12}\rho + {N\over 12}(a-1)\phi +
 e^{-2\phi}\>,\label{vii}
 \ee
and work in the conformal gauge it is equivalent to a Liouville theory
\be
 S={1\over\pi}
 \int d^2x  \left[
 {12\over N}(-\partial_+\chi\partial_-\chi
 +\partial_+\Omega\partial_-\Omega)
 +\lambda^2 e^{{24\over N}
 \left(
 \chi-
 \Omega \right)}	
 + {1\over2} \sum_{i=1}^N \partial_+ f_i \partial_- f_i \right]
 \>.
 \label{viii}
 \ee
The equations of motion of this theory take a very simple form
\be
\partial_+\partial_- (\chi -\Omega)=0 
\>,
 \label{ix}
\ee
\be
\partial_+\partial_- \chi=-\lambda^2 e^{{24\over N}
 \left(
 \chi-
 \Omega \right)}	
 \>.
 \label{x}
 \ee
In addition to these equations the solutions to the equations of motion
have also to satisfy the constraints (that are obtained by variation of 
the full covariant action with respect to $g^{\pm\pm}$)
\be
{N\over 12}t_{\pm}={12\over N}(-\partial_{\pm}\chi\partial_{\pm}\chi
 +\partial_{\pm}\Omega\partial_{\pm}\Omega)+ \partial^2_{\pm}\chi +
{1\over 2} \sum_{i=1}^N \partial_{\pm} f_i \partial_{\pm} f_i 
 \>,
 \label{xi}
 \ee
where $t_{\pm}(x^{\pm})$ are functions of their arguments and depend on 
boundary conditions such as the choice of the quantum state for the 
radiation fields. \\
We can always choose the Kruskal gauge 
$\chi=\Omega$ (i.e. $\rho=\phi$)
for which the general static solutions with no radiation at infinity 
in terms of the original fields read
\be
e^{-2\phi}+{Na\over12}\phi=-\lambda^2x^+x^- -{N\over48}\ln\left(-\lambda^2
x^+x^-\right)+C\> \label{xii}
\ee
and $C$ is an integration constant. We can think of these solutions
as being the ``semiclassical versions'' of those
in (\ref{iv}).  \\
It is easy to to realize that the linear dilaton vacuum (\ref{v}) is not included
in these solutions unless $a={1\over 2}$, in which case $\lambda C$ is identified with
the ADM mass. For different values of $a$ the only solutions that are completely 
regular are those for which
 \be
C= \hat C\equiv -{N\over 48}(1-\ln {N\over48}) + {N a\over 24}(1-\ln {Na\over 24})
\>. \label{xiii}
\ee
The spacetime geometry for these cases is asymptotically minkowskian as $x^+x^-\to\infty$
(and $e^{2\phi}\to 0$).  In the strong-coupling regime the critical line where 
$\Omega^{'}(\phi)=0$, i.e. $e^{-2\phi}={Na\over 24}$, is generically at a finite
distance (see \cite{Cruz.Salas}) except for $a=0$, where it takes the form of 
a semiinfinite throat (we refer to \cite{BPP} for the details). This regular boundary 
can be considered on the same footing as the surface $r=0$ of 4d Minkowski spacetime.\\
Considering the case $C<\hat C$ one can show that we now have a timelike 
curvature singularity
along the line
\be
-\lambda^2x^+x^--{N\over48}\ln\left(-\lambda^2
x^+x^-\right)+C = {N a\over 24}(1-\ln {Na\over 24})
\>. \label{xiv}
\ee
When, instead, $C>\hat C$ the spacetime geometry 
presents light-like weakly coupled
singularities at $x^{\pm}=0$.
It is however consistent in both these cases 
(see \cite{BPP} for $a=0$) to impose reflecting
boundary conditions 
on a suitable timelike hypersurface 
in order to avoid the region of strong coupling in the physical spacetime.
The dynamical evolution of the boundaries for $C<\hat C$ has been considered
in \cite{DAMU,CHUVE} for the RST model and in \cite{BOPP} for the BPP model.\\
The regularity of the solutions with $C= \hat C$ together with the fact that 
they represent the end-point of the Hawking evaporation of these models 
(see \cite {Cruz.Salas}) suggests that  they can be considered the ground state 
of the theory. Any generic solution would therefore have ADM mass 
$\lambda(C-\hat C)$. \footnote{We note, although it could seem superfluous,
 that this definition of mass reduces to the 
classical ADM mass $M$ in the classical limit.} 

\section{Black hole formation with a shock-wave}

We begin our analysis of the dynamical solutions with infalling matter
by recalling that the general solutions to the equations of motion (\ref{ix}),
(\ref{x}) and (\ref{xi}) take the form 
\be
e^{-2\phi}+{Na\over12}\phi=-\lambda^2x^+(x^- + {P(x^+)\over \lambda^2})
-{N\over48}\ln\left(-\lambda^2
x^+x^-\right)+ {M(x^+)\over \lambda} + C
\>, \label{xv}
\ee
where $P(x^+)=\int dx^+ T_{++}^f$ and $M(x^+)=\lambda \int dx^+ x^+T_{++}^f$
are, respectively, the Kruskal momentum and energy of the infalling matter
(and $T_{++}^f= {1\over 2} \sum_{i=1}^N \partial_{+} f_i \partial_{+} f_i$).\\
As a first simple example, let us consider the case of a shock-wave carrying an
energy $m$ and propagating along the null line $x^+=x_0^+$, 
described by the energy-momentum tensor 
\be
T^f_{++}={m\over{\lambda x_0^+}}\delta\left(x^+-x^+_0\right) 
\>. \label{xvi}
\ee
Consider our initial static configuration to be one of those 
studied in the previous section with generic $C$. 
The solution (\ref{xv}) is then 
 \bea
e^{-2\phi}+{Na\over12}\phi
&=&-\lambda^2x^+x^--{N\over48}\ln
\left(-\lambda^2x^+x^-\right)\nonumber\\
&&-{m\over\lambda x^+_0}\left(x^+-x^+_0
\right)\theta\left(x^+-x^+_0\right)+C
\>.\label{xvii}
\eea
The critical line at the future of the shock-wave is therefore given by 
\be
\alpha
=-\lambda^2x^+\left(x^-+\Delta\right)-{N\over48}\ln\left(
-\lambda^2x^+x^-\right)+{m\over\lambda}+C\>,\label{xviii}
\ee
where we have defined 
$\Delta={m\over\lambda^3x_0^+}$ 
and $\alpha={Na\over24}\left(1-\ln
{Na\over24}\right)$.\\ 
Let us first consider the case $C\le \hat C$.  
The onset of the black hole phase is when this curve becomes light-like. We then
expect an apparent horizon to form thus shielding the singularity from the 
external observers and, in the future asymptotic region, the evaporation to
take place as it has been shown in \cite{Cruz.Salas}. 
Once the singularity has become spacelike
it is no more possible to impose reflecting boundary conditions,
which would then violate causality, and the 
spacetime becomes ``truly'' singular.\\
Differentiating eq. (\ref{xviii}) we get 
\be
{dx^-\over{dx^+}}={{\lambda^2(x^- + \Delta)+
{N\over{48x^+}}}\over{-\lambda^2x^+
-{N\over{48x^-}}}}
\>.\label{xix}
\ee
The critical condition 
\be
ds^2= -e^{2\rho_{cr}}dx^+dx^-=0
\>, \label{xx}
\ee
where $\rho_{cr}$ is the value of $\rho$ along the critical line, is then given,
at $x^+=x_0^+$, in terms of a critical mass $m_{cr}$ 
 \be
\lambda^2 (x_0^- + {m_{cr}\over{\lambda^3 x_0^+}})+{N\over{48x_0^+}}=0
\> \label{xxi}
\ee
(here $x_0^-$ is given by eq. (\ref{xviii}) at $x^+=x_0^+$). 
 Combining eqs. (\ref{xviii}) and  (\ref{xxi}) we get    
\be
\alpha - C=({m_{cr}\over\lambda}+{N\over 48})
-{N\over 48}\ln({m_{cr}\over\lambda}+{N\over 48})
\>. \label{xxii}
\ee
In order to understand better this equation, let us first consider 
as our initial configuration the ground state solution 
$C= \hat C = -{N\over 48}(1-\ln {N\over48}) + \alpha$. This gives 
simply $m_{cr}=0$, as already verified in the case of the RST model in \cite{RST}.\\
To analyse the other cases let us write $C=\hat C + \beta$ and define, for
simplicity,
$y\equiv {m_{cr}\over\lambda}$ ($>0$) and $b\equiv {N\over 48}$. 
Eq. (\ref{xxii}) can then be rewritten as 
\be
b\ln (1+{y\over b}) -y=\beta
\>. \label{xxiii}
\ee
The graph of the function $f(y)=b\ln(1+{y\over b}) -y$ is represented
 in Fig. I.\\
As $C<\hat C$, i.e. $\beta<0$, eq. (\ref{xxiii})
has one solution
$y_0>0$. As $\beta\ll 1$ we can expand the logarithm and find
\be
m_{cr}\sim \lambda\sqrt{{{-2N(C-\hat C)}\over 48}}
\>. \label{xxiv}
\ee
The existence of a critical mass \footnote{See, for the BPP model,
\cite{BOPP}.}
can be understood by considering the classical
limit of the solutions (\ref{xii}) with $C<\hat C$, i.e. (\ref{iv}) with $M<0$.
Also in this case there is a critical mass for the formation of the black hole
given by $|M|$. $m_{cr}$ given by (\ref{xxiii}) can then be interpreted as
the
analogous of such a classical critical mass.\\
Turning now to the case $C>\hat C$, i.e. $\beta>0$, we see that eq.
(\ref{xxiii}) has no solution. This is of no surprise,
because the corresponding initial static solution is already a black hole in 
the classical limit! The singularity curve $x^-=0$ persists until the time
$x_1^+=x_0^+(1+ {\lambda\over m}{\beta})$. At this point the singularity
is given by
the critical line (\ref{xviii}), which again becomes light-like at the
end-point
of the evaporation process (Fig. II).
 
\section{Constant energy density flux}

In this section a more general flux of matter will be considered, namely
an influx of constant energy density $\lambda\epsilon$
starting at $x^+=x_0^+$. In the Kruskal gauge it is described by the energy-momentum
tensor 
\be
T^f_{++}(x^+)={\epsilon\over{\lambda(x^+)^2}}\theta(x^+-x_0^+)
\>. \label{xxv}
\ee
The solution to the equations of motion is, from eq. (\ref{xv}),  
\be
e^{-2\phi}+{Na\over12}\phi = -\lambda^2 x^+ (x^- + {\epsilon\over{\lambda^3x_0^+}})
-{N\over48}\ln(-\lambda^2x^+x^-)+{\epsilon\over\lambda}(1+\ln{x^+\over x_0^+})+C
\>. \label{xxvi}
\ee
The analysis of the formation of the black hole proceeds qualitatively
 as in the previous section. The critical line $e^{-2\phi}={Na\over 24}$ 
is now described by the curve 
\be
\alpha= -\lambda^2 x^+ (x^- + {\epsilon\over{\lambda^3x_0^+}})
-{N\over48}\ln(-\lambda^2x^+x^-)+{\epsilon\over\lambda}(1+\ln{x^+\over x_0^+})+C
\> \label{xxvii}
\ee
and the critical condition $ds^2=0$ along this surface defines the 
relation
\be
\lambda^2(x^- +{\epsilon\over{\lambda^3x_0^+}})+ ({N\over 48}-{\epsilon\over\lambda})
{1\over x^+}=0.
\> \label{xxviii}
\ee 
We can now rewrite (\ref{xxvii}) using (\ref{xxviii}) in the form 
\be
\alpha - C= {N\over 48}(1-\ln[({N\over 48}-{\epsilon\over\lambda})
+{{\epsilon x^+}\over{\lambda x_0^+}}]) +
{\epsilon\over\lambda}
\ln{x^+\over{x_0^+}}
\>.\label{xxix}
\ee 
The case of the ground state solution $C=\hat C$ requires 
\be
{\epsilon\over\lambda}={\epsilon_{cr}\over\lambda}\equiv {N\over 48}
\>.\label{xxx}
\ee
Provided we introduce the asymptotic minkowskian null coordinate
$\sigma^+={1\over\lambda}\ln\lambda x^+$ we find that the threshold for 
black hole formation is given by an energy flux of the form
\be
T^f_{\sigma^+\sigma^+}={{N\lambda^2}\over{48}}
\>.\label{xxxi}
\ee
This is nothing but the rate of evaporation of these two-dimensional black 
holes (see for instance \cite{CGHS}) and this result is quite plausible
 because we wouldn't expect, on physical grounds, a black hole to form for subcritical
fluxes because of the semiclassical Hawking effect. The same result was obtained 
in the RST model in \cite{RST}.\\
To analyse eq. (\ref{xxix}) for other values of $C$ we introduce, in order to
simplify the expression, the quantities $x\equiv{x^+\over{x_0^+}}$ ($>1$),
$\epsilon$ instead of ${\epsilon\over\lambda}$, $b\equiv {N\over 48}$
and $\beta=C-\hat C$. We then rewrite (\ref{xxix}) as 
\be
b\ln[{\epsilon\over b} x + 
(1-{\epsilon\over b})] -\epsilon\ln x=\beta
 \>.\label{xxxii}
\ee
On the basis of the results for the case $C=\hat C$ ($\beta=0$) we will
consider the ``subcritical'' $\epsilon<b$ and ``supercritical'' $\epsilon>b$
fluxes separately. \\
For $\epsilon<b$ the graph of the function 
$g(x)=b\ln[{\epsilon\over b}x + 
(1-{\epsilon\over b})] -\epsilon\ln x$
is represented in Fig. III. We see that as $\beta<0$ eq. (\ref{xxxii}) is
never
satisfied, which means that with this subcritical flux the black hole is never
formed, in complete analogy with the case $\beta=0$. 
Turning to $\beta>0$ we find a rather surprising result: for any values of 
$\epsilon<N/48$ Hawking radiation is always produced!. 
The singularity curve $x^-=0$  transforms into a space-like curve at the time 
given by the condition
\be
 {x_1^+}-{x_0^+}\ln{x_1^+\over{x_0^+}}= x_0^+ (1+ {\lambda\over
{\epsilon}}{\beta})
 \> \label{xxxiii}
 \ee
and finally it turns out to be light-like at the end-point of Hawking 
evaporation $x_2^+\equiv x_0^+g^{-1}(\beta)$ (see again Fig.II).

The supercritical flux $\epsilon >b$ gives the function $g(x)$ in Fig.
IV.
We see clearly that as $\beta<0$ the black hole forms at the time $x_0^+g^{-1}(\beta)$
( $>x_0^+$). In this case there is a critical mass given by
$\epsilon \ln g^{-1}(\beta)$. 
This happened also 
in the shock-wave scenario analysed in the previous section and 
the  possible physical interpretation is therefore the same. 
On the other hand, for $\beta>0$ the eq. $g(x)=\beta$ has no real solution, i.e. the black
hole starts to radiate but never disappears.
\bigskip

\section{Discussion and conclusions}
We could ask, at this point, whether the results obtained in the last two sections
are only specific to the types of infalling matter considered.
We can show quite easily that they have instead a rather general validity.
In the general case, for infalling fluxes of matter switched on at $x^+=x_0^+$,
the critical line is given by
\be
\alpha=-\lambda^2x^+(x^- + {P(x^+)\over \lambda^2})
-{N\over48}\ln\left(-\lambda^2
x^+x^-\right)+ {M(x^+)\over \lambda} + C
 \>. \label{xxxiv}
\ee
The time at which the singularity becomes null is related to the critical Kruskal
momentum $P_{cr}(x^+)$ through the equation
\be
\lambda^2(x^- +{{P_{cr}(x^+)}\over{\lambda^2}}) + {N\over{48 x^+}}=0
 \>. \label{xxxv}
\ee
Combining the previous two equations and considering the quantities $\beta$ and
$b$ defined in the last section we obtain
\be
\beta=b\ln[1+{{x^+P_{cr}(x^+)}\over b}] -{M_{cr}(x^+)\over\lambda}
 \>. \label{xxxvi}
\ee
The function 
$h(x^+)=b\ln[1+{{x^+P_{cr}(x^+)}\over b}] -{M_{cr}(x^+)\over\lambda}$
is the analogue of $f(y)$ and $g(x)$ considered in sections 4 and 5. \\
Starting from $h(x_0^+)=0$ the behaviour of this function 
for $x^+>x_0^+$ is essentially
given by its first derivative
\be
h^{'}(x^+)={{P_{cr}(x^+)(1-{{x^{+2}T_{++}^{cr}}\over b})}
\over{1+{{x^+P_{cr}(x^+)}\over b}}}
 \>. \label{xxxvii}
\ee
Provided that $P_{cr}(x^+)>0$ (which is always true for classical matter) we 
easily see that $h^{'}(x^+)<0$ for $T_{++}^{cr}>{N\over{48x^{+2}}}$ 
and $h^{'}(x^+)>0$ as $T_{++}^{cr}<{N\over{48x^{+2}}}$. The qualitative
behaviour of
the function $h(x)$ is therefore the same as in Figs. III and IV.
\\ \\ 
We can now summarize the results of our investigation as follows.
We have considered initial static geometries parametrized by the 
continuous parameter $C$.
 As $C\leq\hat C$, where $\hat C$ denotes the ground
state solution, there is essentially a threshold on the energy density 
of the incoming radiation $\epsilon_{cr}={{N\lambda^2}\over 48}$ given by
the Hawking rate of evaporation. For $\epsilon<\epsilon_{cr}$, in fact, 
it is not possible to form the black hole and as $\epsilon>\epsilon_{cr}$ 
 there is, in addition, also a critical mass 
(vanishing when $C=\hat C$). 
When $C>\hat C$ the static semiclassical solution 
can be interpreted as a sort of ``black hole'' in an (unstable)
equilibrium state.
By sending in a small amount of energy
one induces the evaporation process, irrespective of the incoming
density flux $\epsilon$ and with no critical mass. This is in
contrast with the thermal equilibrium black hole solutions which maintain the
equilibrium even in the presence of incoming matter.\\

We would like to mention that it could be of interest to study the critical
behaviour for black hole formation in other solvable models of 2d dilaton gravity
with a different thermodynamic \cite{Cruz.Salas2}. This will be considered
in a future publication.

\section*{Acknowledgements}
J. N-S would like to thank J. Cruz for collaboration in early states of this 
work.

\begin{figure}
\centerline{\epsfxsize=10cm \epsfbox{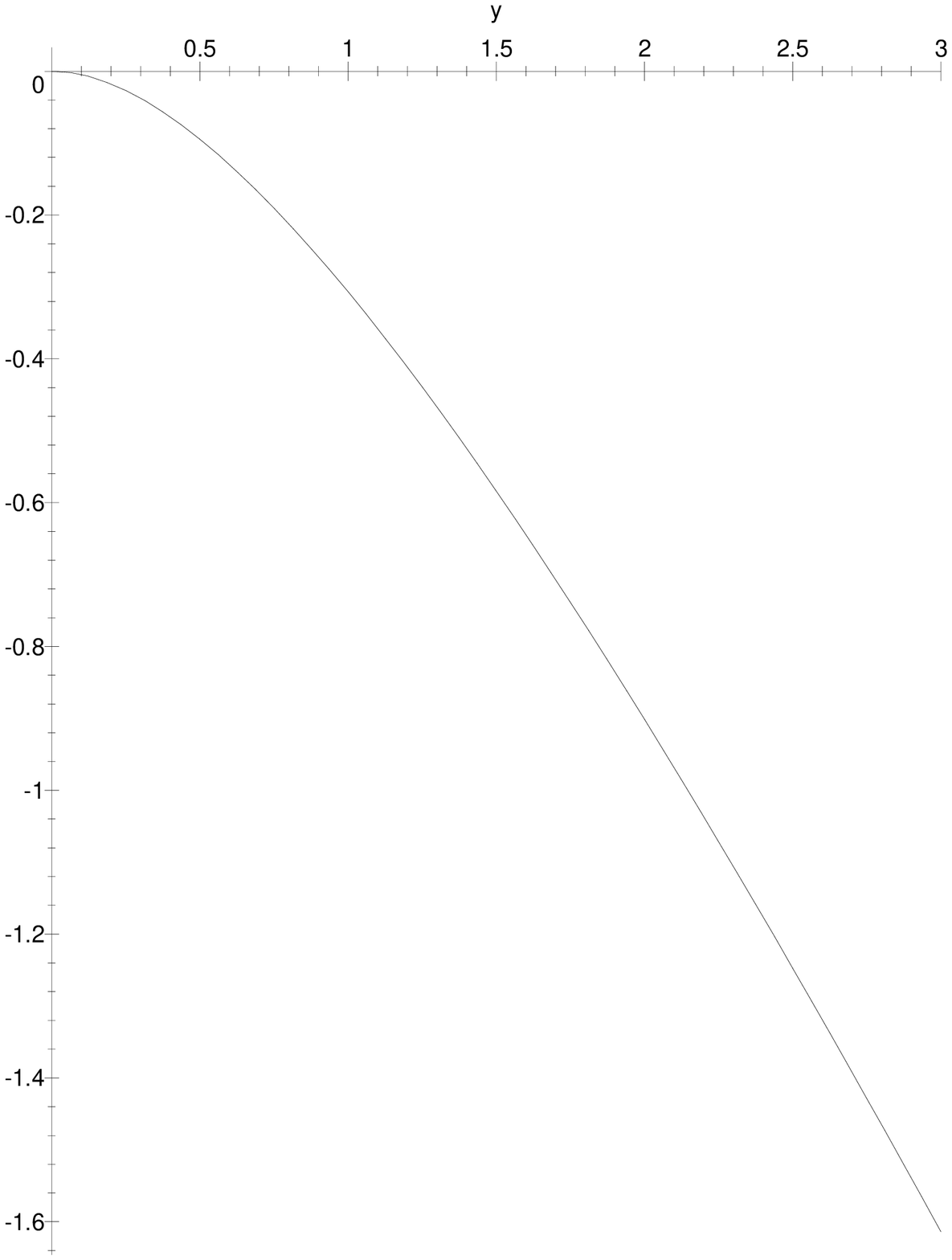}}
\bigskip
\caption{Graph of the function $f(y)$ (we chose $b=1$).}
\end{figure}

\begin{figure}
\centerline{\epsfysize=12cm \epsfbox{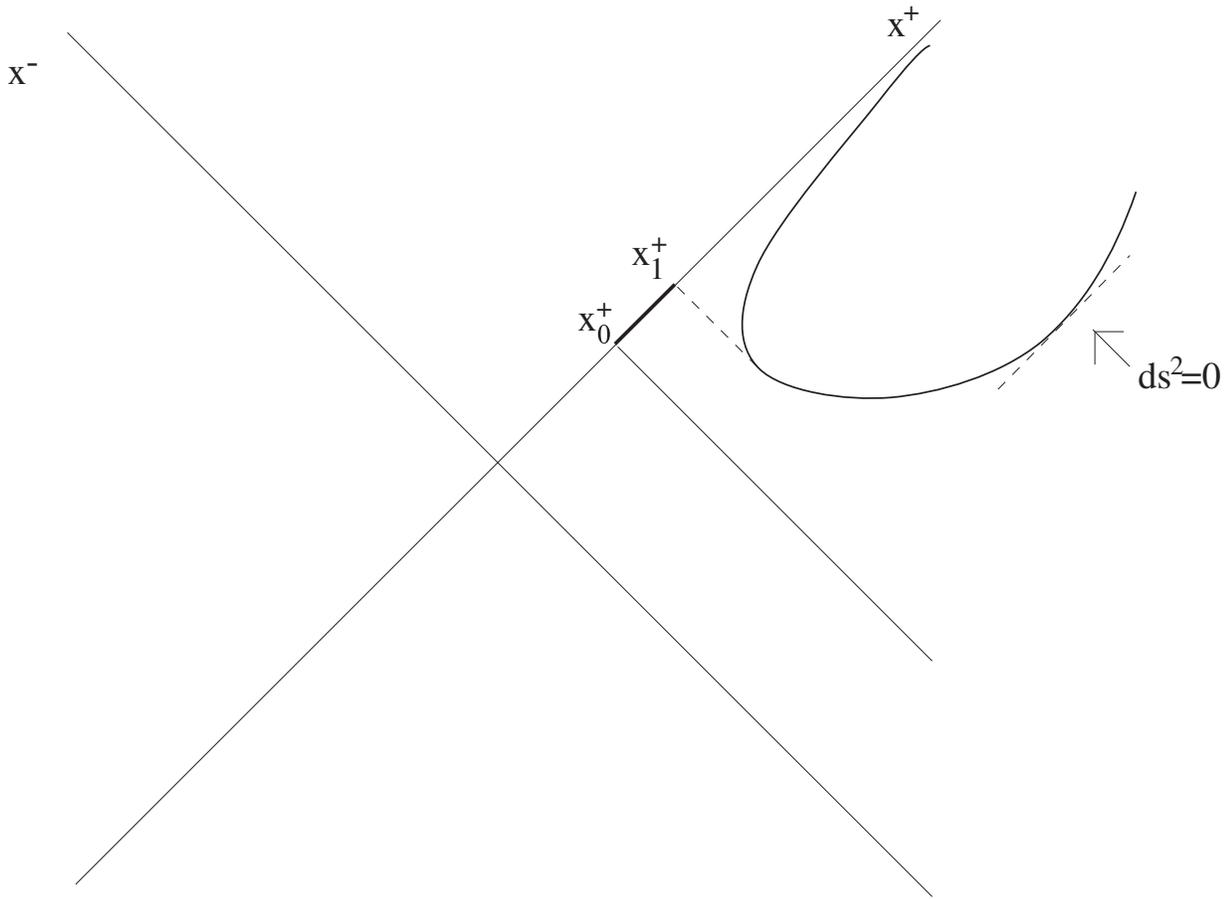}}
\bigskip
\caption{Behaviour of the singularity curve for $\beta>0$ both for the
shock-wave and $\epsilon<N/48$ cases.}
\end{figure}  

\begin{figure}
\centerline{\epsfysize=7cm \epsfbox{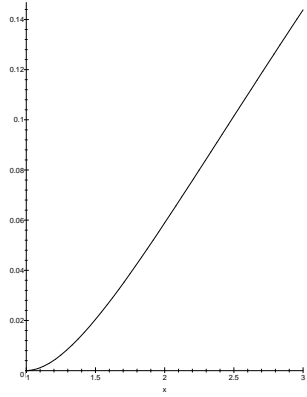}}
\bigskip
\caption{$g(x)=b\ln [{\epsilon\over b}x + (1-{\epsilon\over b})] -
\epsilon\ln x$ for $b=1$ and $\epsilon=1/2$ (subcritical flux).}
\end{figure} 

\begin{figure}
\centerline{\epsfysize=7cm \epsfbox{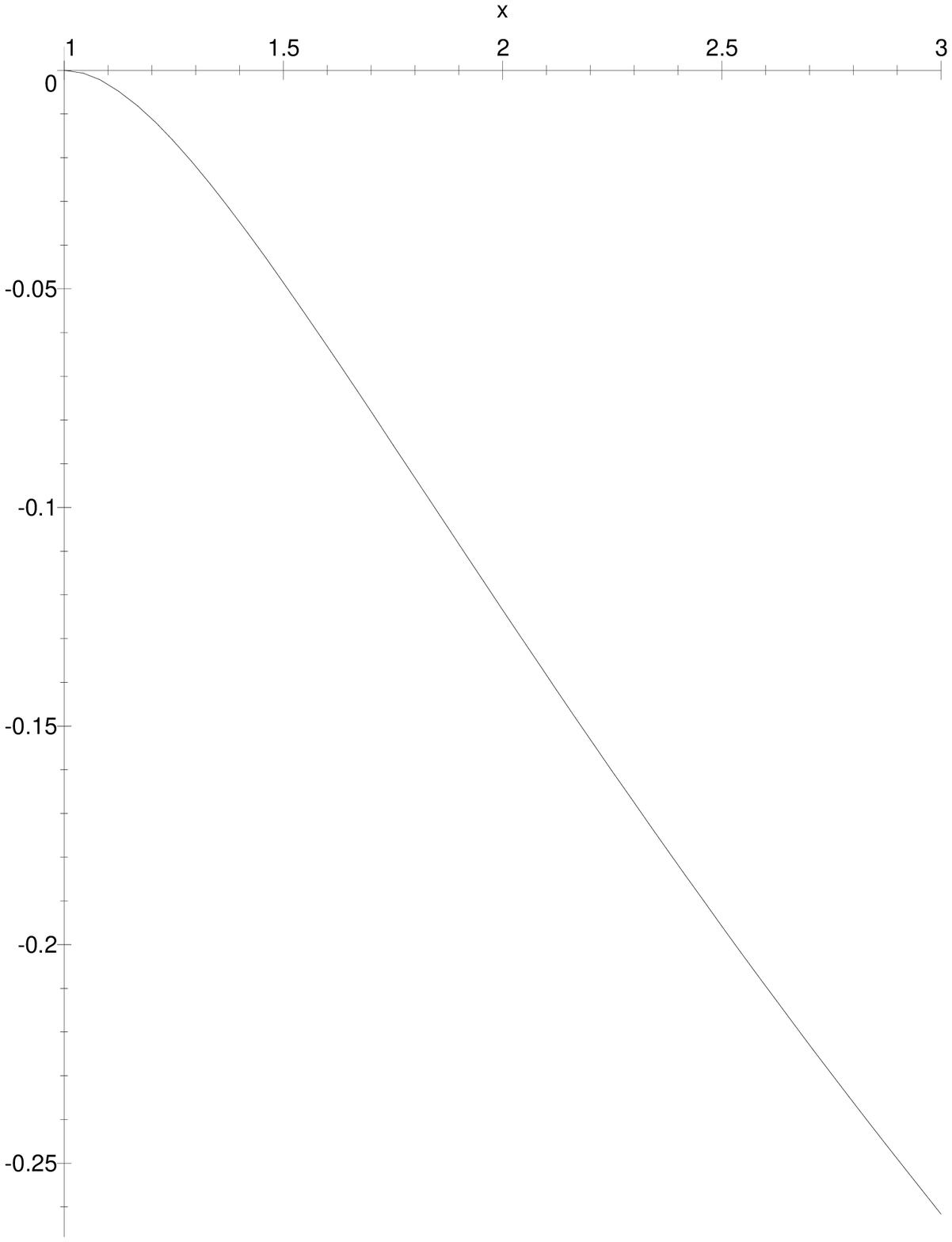}}
\bigskip
\caption{$g(x)$ as in Fig. III, but with $b=1$ and $\epsilon=3/2$
(supercritical flux).}
\end{figure}


\begin{thebibliography}{99}


\bibitem{Hawking}S. W. Hawking, Commun. Math. Phys. 43 (1975) 199

\bibitem{Bilal.Callan}A.  Bilal and C.  Callan, Nucl. Phys. B 394 (1993) 73 ;
S.  de Alwis, Phys. Lett. B 289 (1992) 278

\bibitem{Cruz.Salas}J. Cruz and J. Navarro-Salas, Phys. Lett. B 375 (1996) 47 

\bibitem{Choptuik}M. W. Choptuik, Phys. Rev. Lett. 70 (1993) 9

\bibitem{RST}J. G. Russo, L.  Susskind and L.  Thorlacius, Phys.  Rev. D
46
(1993) 3444 ; D 47 (1993) 533

\bibitem{BPP}S.  Bose, L.  Parker and Y.  Peleg, Phys. Rev. D 52 (1995) 3512

\bibitem{CGHS}C.  G.  Callan, S.  B.  Giddings, J.  A.  Harvey
and A.  Strominger, Phys. Rev. D 45 (1992) 1005

\bibitem{DAMU}S. R. Das and S. Mukherji, Mod. Phys. Lett. A 9 (1994) 3105 ;
Phys. Rev. D 50 (1994), 930  

\bibitem{CHUVE}T. D. Chung and H. Verlinde, Nucl. Phys. B 418 (1994) 305 


\bibitem{BOPP}S. Bose, L. Parker and Y. Peleg, Phys. Rev. Lett. 76 (1996), 861 


\bibitem{Cruz.Salas2} J. Cruz and J. Navarro-Salas, Phys. Lett. B 387 (1996) 51

\end{thebibliography}
 \end{document}